\documentclass{iopjournal}

\begin{document}


\title{Fidelity bounds for spin-dependent kicks with pulsed lasers}

\author{C. Sagaseta$^{1,*}$\orcid{0009-0007-4231-1201}, H. Liu$^2$\orcid{0000-0001-8924-2900}, V. D. Vaidya$^2$\orcid{0000-0002-0586-8568}, C. R. Viteri$^2$\orcid{0009-0009-7482-2430}, J. J. Garc{\'\i}a-Ripoll$^3$\orcid{0000-0001-8993-4624} and E. Torrontegui$^{1,*}$\orcid{0000-0002-9148-4113}}

\affil{$^1$Department of Physics, Universidad Carlos III de Madrid, Avenida de la Universidad 30, 28911 Legan\'es, Madrid, Spain}

\affil{$^2$IonQ, Inc., 4505 Campus Drive, College Park, MD 20740, USA}

\affil{$^3$Instituto de F\'{\i}sica Fundamental, Consejo Superior de Investigaciones Cient{\'\i}ficas (IFF-CSIC), Calle Serrano 113b, 28006 Madrid, Spain}

\affil{$^*$Authors to whom any correspondence should be addressed.}

\email{csagaset@fis.uc3m.es, eriktorrontegui@gmail.com}


\begin{abstract}
Excitation of trapped-ion hyperfine qubits with fast optical Raman pulses enables faster-than-trap-period entangling gates with qubits of long coherence time for practical quantum computation. Achieving high-fidelity fast two-qubit gates requires high-quality spin-dependent kicks (SDKs), which form their fundamental building blocks. Here, we characterize the control parameters (including Raman frequency difference, pulse arrival times, Lamb--Dicke parameter, temperature, pulse width, and SDK time) that maximize the performance of single-ion SDKs for protocols compatible with performed experiments involving a small number of fast pulses. We demonstrate through analytical methods and numerical simulations that, within the model commonly used for infidelity optimization, finite pulse duration is the dominant source of error, exceeding the contribution of secular motion by orders of magnitude for nanosecond-scale SDKs. Low infidelities---below $10^{-3}$ for schemes with $\gtrsim\!10$ fixed-amplitude, equispaced, picosecond pulses---are achievable in SDK times on the order of nanoseconds. These results provide quantitative design rules for achieving competitive SDK fidelities with current pulsed-laser technology, laying the foundation for sub-microsecond trapped-ion quantum entangling operations.
\end{abstract}

\section{Introduction}

Trapped ions are a leading platform for quantum information processing~\cite{bruzewicz2019}. Qubits encoded in internal electronic states exhibit extremely long coherence times~\cite{wang2021,harty2014} and can be manipulated with high precision using electromagnetic fields \cite{blatt2008}. Among all quantum computing platforms, trapped ions currently achieve record fidelities for state preparation and measurement \cite{an2022}, as well as for single-qubit~\cite{smith2025} and two-qubit gates, both electronic and laser-based~\cite{loschnauer2025,clark2021}. These capabilities have enabled demonstrations of quantum error correction~\cite{zhang2020,egan2021,ryan-anderson2021,nguyen2021,postler2024} and the implementation of small-scale quantum algorithms~\cite{monz2016,debnath2016,hempel2018,liu2025,figgatt2017}.

Most quantum gates in trapped-ion systems operate on adiabatic timescales with respect to the motional dynamics of the ions, since they rely on spectroscopically resolving individual modes of motion~\cite{molmer1999}. As the number of ions increases, gate operations become more difficult to realize due to normal-mode crowding and unwanted couplings \cite{leung2018,mai2025,choi2014}, effectively imposing a speed threshold for achieving high fidelity.

In terms of scalability, the dominant architecture is the quantum charge-coupled device (QCCD) ~\cite{kielpinski2002,akhtar2023}, which combines small ion registers with ion transport between functional zones. This scheme allows the record fidelities achieved in the smallest processors to be approached while scaling system size~\cite{moses2023}. Connectivity can be provided either by shuttling ions between devices~\cite{akhtar2023,sterk2022} or by linking them through photonic interconnects~\cite{kaushal2019,main2025,monroe2014,knollmann2024,oreilly2024,stephenson2020,drmota2023}. However, in such modular architectures the dominant speed bottleneck arises from ion transport, as the ions must be moved and subsequently recooled before gate operations can be performed~\cite{pino2021}.

Fast gates provide an alternative approach to overcoming the scalability--speed trade-off while maintaining the required high fidelity~\cite{garcia-ripoll2003,garcia-ripoll2005,duan2004}. In these schemes, broadband laser pulses impulsively excite the ions to generate spin--motion entanglement without requiring cooling to the Lamb--Dicke regime~\cite{mizrahi2013,johnson2015}, enabling two-qubit gates on timescales faster than the motional periods of the ions~\cite{mizrahi2014}. Theoretical studies further show that fast two-qubit gates can be performed with high fidelity between arbitrary pairs of ions in a long chain~\cite{bentley2015,ratcliffe2018,mehdi2021,mehdi2020,savill-brown2025a,savill-brown2025}, potentially reducing resource overhead in modular architectures implementing fast gates.

These fast gates are engineered through sequences of spin-dependent kicks (SDKs)~\cite{mizrahi2014,wong-campos2017}, which couple the ion spin and momentum degrees of freedom: a spin flip is accompanied by a momentum change whose sign depends on the spin state. The performance of fast two-qubit gates depends directly on the quality of individual SDKs, though agnostic to the specific SDK implementation, so achieving high-fidelity kicks is paramount~\cite{savill-brown2025}. Precursor studies established the underlying mechanism and the ideal conditions required to generate a SDK using pulsed lasers~\cite{mizrahi2013,mizrahi2014}. While fast single-qubit~\cite{campbell2010,putnam2024} and two-qubit gates~\cite{wong-campos2017} have been demonstrated, the parameter regimes needed to create SDKs with pulsed protocols have been derived under three approximations---high-pulse number trains, instantaneous pulses, and negligible ion motion~\cite{mizrahi2013,mizrahi2014}--- which limit the fidelities achievable under realistic experimental conditions~\cite{wong-campos2017}.

For improved quantum computing performance, it is desirable to have protocols with a small number of pulses~\cite{mizrahi2013,wong-campos2017,johnson2017} to shorten SDK times and minimize the impact of low-frequency noise. This motivates studying the optimal conditions for few-pulse protocols and characterizing the errors that arise when finite pulse width and ion motion are included in the model, thereby providing more realistic fidelity limits. To address these questions, we follow a bottom-up approach, where we first identify the optimal parameters in the idealized model for a small number of pulses and then sequentially relax each approximation to study their impact on the infidelity.

This work achieves theoretical characterization of all relevant aspects that influence on SDK performance, namely Raman frequency difference, repetition time between pulses, Lamb--Dicke parameter, temperature, pulse width, and SDK time. We show that low infidelity---below $10^{-3}$ for protocols with $\gtrsim\!10$ fixed-amplitude, equally-spaced, picosecond pulses---is attainable in SDK times of a few nanoseconds. A longer pulse width significantly impacts the SDK protocol, highlighting the importance of considering it in parameter optimization. In contrast, we find that ion secular motion has a small effect during the few-nanosecond-long SDK, with an error of the order of $10^{-5}$.

The manuscript is organized as follows. Sec.~\ref{sec:model} presents the model and the pulsed protocol used to create a SDK. Sec.~\ref{sec:delta pulses} deals with protocols composed of a small number of delta pulses exciting a non-moving ion, and the optimization of parameters to maximize the SDK process. In Sec.~\ref{sec:finite pulses} we find the error caused by considering the finite duration of the pulses, while in Sec.~\ref{sec:ion motion} we analyze the impact of the ion motion on the SDK protocol. Conclusions and outlook are given in Sec.~\ref{sec:conclusions}.

\section{Model}\label{sec:model}

\subsection{Raman-illuminated trapped ion}

We consider a single ion in a harmonic potential interacting with two Raman beams. The qubit is encoded in the ground-state hyperfine levels and the Raman beams arrive from counter-propagating directions along $x$ with orthogonal linear polarizations $(\text{lin}\perp\text{lin})$, both orthogonal to an applied static magnetic field defining the quantization axis; see Fig.~\ref{fig1}(a). This polarization configuration produces a polarization gradient at the position of the ion, which couples spin and motion~\cite{metcalf1999}. Assuming that the pulses from each beam have the same envelope and they arrive simultaneously at the ion, the effective Hamiltonian in the RWA after adiabatically eliminating the excited states is given by $(\hbar=1)$~\cite{mizrahi2013,mizrahi2014}
\begin{equation}\label{eq: H}
\hat{H}=\underbrace{\omega_\text{t}\hat{a}^\dagger\hat{a}\vphantom{\frac{a}{b}}}_{\displaystyle\hat{H}_\text{t}} + \underbrace{\frac{\omega_\text{hf}}{2}\hat{\sigma}_z}_{\displaystyle\hat{H}_\text{hf}} + \underbrace{\mathcal{W}(t)\sin\big(k_\text{d}\hat{x} + \delta t+\varphi_0\big)\hat{\sigma}_x\vphantom{\frac{a}{b}}}_{\displaystyle\hat{H}_\text{int}(t)},
\end{equation}
where $\hat{a}^\dagger$ ($\hat{a}$) is the creation (annihilation) operator of the motional mode, $\hat{\sigma}_x$ and $\hat{\sigma}_z$ are the Pauli matrices, $\omega_\text{t}$ is the trap frequency, $\omega_\text{hf}$ is the hyperfine qubit frequency, $\mathcal{W}(t)$ is the beam envelope, $k_\text{d}=k_1-k_2$ is the difference wavenumber of the counter-propagating beams, $\delta$ is the frequency difference between the beams and $\varphi_0$ is an initial phase. Equation~(\ref{eq: H}) is valid as long as $\Delta\gg\omega_\text{hf}$ (see Fig.~\ref{fig1}(b)) and $\tau\partial_t\mathcal{W}(t)\ll\Delta$, where $\tau$ is the pulse duration. Also, the bandwidth of $\mathcal{W}(t)$ must be much larger than $\omega_\text{hf}$ to be able to resolve the qubit splitting, but much smaller than $\Delta$ to avoid exciting any higher level. We describe the state of the ion by the computational basis of the qubit $\{\ket{Q}\}$, with $Q\in\{0,\,1\}$, times the Fock basis $\{\ket{f}\}$, with $f\in\mathbb{N}_0$.

\subsection{Spin-dependent kicks}\label{sec:SDKs}

\begin{figure}[t!]
\centering
\includegraphics[width=0.6\textwidth]{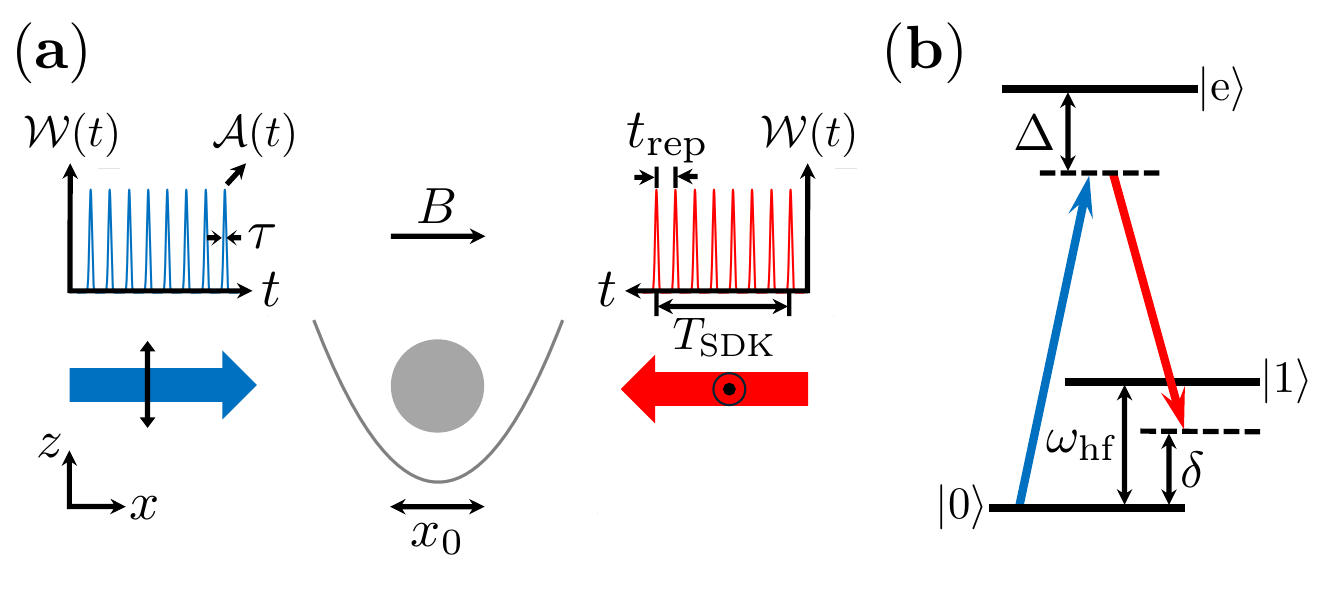}
\caption{Schematic of the pulsed Raman SDK. Two counter-propagating beams in the $\text{lin}\perp\text{lin}$ polarization configuration interact with a single ion (grey ball) confined in a harmonic potential (grey parabola) with characteristic length $x_0$, that corresponds to the zero-point spread of the motional wavepacket of the ion. The quantization axis is defined by the magnetic field $\mathbf{B}=B\mathbf{x}$, with $\mathbf{x}$ the unit vector along $x$. A SDK is generated by two pulse trains $\mathcal{W}(t)$ of duration $T_\text{SDK}$ composed of pulses $\mathcal{A}(t)$ of duration $\tau$ separated by equispaced intervals $t_\text{rep}$. (b) Generic energy level diagram (not to scale) of the Raman transition, where the Raman beams have a frequency difference $\delta$ that does not necessarily match the qubit splitting $\omega_\text{hf}$. The higher-frequency beam is detuned from the excited state $\ket{\text{e}}$ by $\Delta$.}  
\label{fig1}
\end{figure}
We follow the scheme with counter-propagating trains of short pulses, as depicted in Fig.~\ref{fig1}(a), proposed by Mizrahi \textit{et al.}~\cite{mizrahi2013,mizrahi2014} to produce a SDK. The requirements for achieving this were derived taking three approximations: (i) the number of pulses is large, $N\rightarrow\infty$; (ii) the pulse duration $\tau$ is much shorter than the intrinsic qubit dynamics, $2\pi/\tau\gg\omega_\text{hf}\gg\omega_\text{t}$, so the pulses are approximated by a delta function; and (iii) the SDK protocol time $T_\text{SDK}$ satisfies $T_\text{SDK}\ll2\pi/\omega_\text{t}$, so the ion is effectively frozen during the interaction dynamics and motion can be neglected.

Protocols with a large number of pulses are usually impractical, as they would prolong the SDK time, unlike experiments that employ a limited number of pulses~\cite{mizrahi2013,wong-campos2017,johnson2017}. Moreover, in such experiments the pulse duration is not well separated from the qubit dynamical timescale, so that non-negligible qubit evolution may occur during the pulse interaction, making the instantaneous-pulse assumption not sufficient. In addition, the ion undergoes harmonic motion during the SDK time, which can be comparable to the motional period, and thus deteriorate the performance of the SDK protocol. In the following three sections, we eliminate each one of these approximations individually. However, let us now discuss the results given these simplifications.

Mathematically, the target SDK operator takes the form
\begin{equation}\label{eq: U_SDK}
    \hat{U}_\text{SDK} = \hat{\mathcal{D}}(i\eta)\hat{\sigma}^++\hat{\mathcal{D}}(-i\eta)\hat{\sigma}^-,
\end{equation}
where $\hat{\mathcal{D}}(\alpha)=\exp\big\{\alpha\hat{a}^\dagger-\alpha^*\hat
a\big\}$ is the displacement operator, $\hat{\sigma}^+$ $\left(\hat{\sigma}^-\right)$ is the spin raising (lowering) operator, and $\eta=k_\text{d}x_0$ is the Lamb--Dicke parameter, with $x_0=\sqrt{\hbar/\left(2M\omega_\text{t}\right)}$ corresponding to the zero-point motion spread of the ion with mass $M$. If the ion is initially in a state $\ket{Q,\alpha}$, where $\ket{\alpha}$ is a coherent state with amplitude $\alpha\in\mathbb{C}$, the ideal target state produced by the operator in Eq.~(\ref{eq: U_SDK}) is
\begin{equation}\label{eq: target state}
     \hat{U}_\text{SDK}\ket{Q,\alpha}=\ket{Q\oplus1,\alpha+(-1)^Qi\eta},
\end{equation}
where $\oplus$ denotes addition modulo 2.
In our scheme, the Raman beams consist of trains of short pulses, so the interaction term in Eq.~(\ref{eq: H}) becomes
\begin{equation}\label{eq: H_int delta}
    \hat{H}_\text{int}(t)=\sum_{n=0}^{N-1}\hat{H}_\text{pulse}(t,\,t_n),
\end{equation}
with $N$ the number of pulses and the $n$-th pulse pair term given by
\begin{equation}\label{eq: H_p operator}
    \hat{H}_\text{pulse}(t,\,t_n)=\mathcal{A}(t-t_n)\sin\left(\eta\big(\hat{a}^\dagger+\hat{a}\big) + \delta t\right)\hat{\sigma}_x,
\end{equation}
where $\mathcal{A}(t)$ is the pulse envelope and $t_n$ is the arrival time of the $n$-th pulse pair. We have assumed $\varphi_0=0$ without loss of generality.

In the limit given by approximation (ii): $\mathcal{A}(t)=\theta\delta(t)$, with pulse amplitude $\theta=\pi/N$, the evolution operator for a single pulse pair arriving at time $t=t_n$, $\hat{U}(t_n)$, has an analytical expression, so its action on a state $\ket{Q,\alpha}$ results in
\begin{equation}\label{eq:superposition state}
    \hat{U}(t_n)\ket{Q,\alpha}=\sum_{m=-\infty}^\infty e^{-im\delta t_n}J_m(\theta)\hat{\sigma}^m_x\ket{Q,\alpha - im\eta},
\end{equation}
where $J_m$ is the $m$-th order Bessel function of the first kind. The state in Eq.~(\ref{eq:superposition state}) represents an infinite superposition of displaced coherent states with even/odd spin flips, with rapidly decreasing amplitudes when the order $m$ increases. The operator $\hat{U}(t_n)$ by itself is not useful for quantum information applications; however, a suitable combination of these idealized pulses approximates the SDK operator in Eq.~(\ref{eq: U_SDK}) by meeting the conditions~\cite{mizrahi2013,mizrahi2014} 
\begin{align}
    \label{eq: Mizrahi condition 1}
    (\delta-\omega_\text{hf})\,t_n&\ne0 \pmod{2\pi},\\
    \label{eq: Mizrahi condition 2}
    (\delta+\omega_\text{hf})\,t_n&=0 \pmod{2\pi},
\end{align}
that become exact in the limit $N\rightarrow\infty$. The pulse pair arrival times need not be equispaced; however, we consider evenly spaced pulses throughout this work, which is easier to implement experimentally by using ad hoc laser sources~\cite{hussain2021}.

\section{Optimal few-delta-pulse protocols}\label{sec:delta pulses}

In pursuit of faster quantum gates, SDK protocols with a small number of pulses are desirable~\cite{mizrahi2013,wong-campos2017,johnson2017}. Accordingly, in this section we generalize previous SDK theory by removing the infinite-pulse number approximation and instead consider protocols made up of a finite number of delta pulses, while neglecting ion motion, highlighting the differences with the infinite pulse number limit. Under these assumptions, we simplify the description of the model by taking $x$ as a constant of motion and optimize the pulse repetition time $t_\text{rep}$ and Raman frequency difference $\delta$ to best approximate a SDK.

\subsection{Formulation}

We consider the protocol for creating a SDK described in Sec.~\ref{sec:SDKs}. Since ion motion is neglected, $\hat{H}_\text{t}=0$ in Eq.~(1), the position $x$ can be treated as a parameter. Thus, the pulse pair term in Eq.~(\ref{eq: H_p operator}) becomes
\begin{equation}\label{eq: H_p parameter}
    \hat{H}_\text{pulse}(t,\,t_n,\,\phi_{kx})=\mathcal{A}(t-t_n)\sin\big(\phi_{kx} + \delta t\big)\hat{\sigma}_x,
\end{equation}
where $\phi_{kx}=k_\text{d} x=\eta u$, with $u=x/x_0$ the normalized position. The $N$ delta pulses, $\mathcal{A}(t) = \theta\delta(t)$, have the same area $\theta=\pi/N$ and equispaced arrival times $t_n=nt_\text{rep}$, with $n\in\mathbb{N}_0$. The pulse train operator only acts on spin space and is $x$-dependent. It is obtained by concatenation of a pulse pair operator at time $t_n$
\begin{equation}\label{eq: U_tn sc delta}
    \hat{\mathcal{U}}^{(1)}(t_n,\,\phi_{kx})=e^{-i\theta\sin\left(\phi_{kx}+\delta t_n\right)\hat{\sigma}_x},
\end{equation}
and free evolution operator
\begin{equation}\label{eq: U_FEq}
    \hat{\mathcal{U}}_\text{FE}^\text{(hf)}(\mathcal{T})=e^{-i\frac{\omega_\text{hf}}{2}\mathcal{T}\hat{\sigma}_z},
\end{equation}
for time $\mathcal{T}=t_\text{rep}$ to yield
\begin{equation}\label{eq: U_N delta}
    \hat{\mathcal{U}}_N^{(1)}(\phi_\delta,\,\phi_\text{hf,\,}\phi_{kx})= \prod_{n=0}^{N-2}\left[e^{-i\theta\sin\left(\phi_{kx} + (N-n-1) \phi_\delta\right)\hat{\sigma}_x}e^{-i\frac{\phi_\text{hf}}{2}\hat{\sigma}_z}\right]\hat{\mathcal{U}}^{(1)}(t_0,\,\phi_{kx}),
\end{equation}
with the angles defined as $\phi_\delta=\delta t_\text{rep}$ and $\phi_\text{hf}=\omega_\text{hf}t_\text{rep}$.

The SDK target operator in Eq.~(\ref{eq: U_SDK}), with the approximation $\eta\big(\hat{a}^\dagger+\hat{a}\big)\approx\phi_{kx}$, is reexpressed as
\begin{equation}\label{eq: U_SDK sc}
    \hat{\mathcal{U}}_\text{SDK}(\phi_{kx},\,\xi) = e^{i\xi}e^{i\phi_{kx}}\hat{\sigma}^++e^{-i\phi_{kx}}\hat{\sigma}^-
    =\begin{pmatrix}
        0 & e^{i\xi}e^{i\phi_{kx}}\\
        e^{-i\phi_{kx}} & 0
    \end{pmatrix},
\end{equation}
where we have introduced $\xi$ as the relative phase between upward $\big(\text{positive},\,e^{i\phi_{kx}}\big)$ and downward $\big(\text{negative},\,e^{-i\phi_{kx}}\big)$ kick (phase), which we need to incorporate in the target operator so that $\hat{\mathcal{U}}_N^{(1)}$ and $\hat{\mathcal{U}}_\text{SDK}$ can be faithfully compared. We use the state-averaged process fidelity as the figure of merit to measure how close the pulse train operator in Eq.~(\ref{eq: U_N delta}) is to the SDK operator in Eq.~(\ref{eq: U_SDK sc})
\begin{equation}\label{eq: F sc}
    \bar{\mathcal{F}}(\phi_\delta,\,\phi_\text{hf})=\underset{\xi \in [0,\,2\pi)}{\max}\frac{1}{2^2}\int_{-\infty}^\infty\left|\sum_{Q}\bra{Q}\hat{\mathcal{U}}_\text{SDK}^\dagger(u,\,\xi)\,\hat{\mathcal{U}}_N^{(1)}(u)\ket{Q}\right|^2P_{\hat{\rho}}(u)du,
\end{equation}
where $P_{\hat{\rho}}(u)=\bra{u}\hat{\rho}\ket{u}$ is the probability that the ion is at position $u$ assuming it is in a motional state $\hat{\rho}=\ket{\alpha}\bra{\alpha}$ (we generalize Eq.~(\ref{eq: F sc}) by introducing $\hat{\rho}$, as we will also consider thermal states). Equation~(\ref{eq: F sc}) is interpreted as the process fidelity at every position weighted by the motional wavefunction of the ion. For a more meaningful discussion of the results, in what follows we will use the state-averaged infidelity
\begin{equation}\label{eq: I_sc}
    \bar{\mathcal{I}}=1-\bar{\mathcal{F}}.
\end{equation}

\subsection{Methodology}

\begin{figure}[t!]
\centering
 \includegraphics[width=\textwidth]{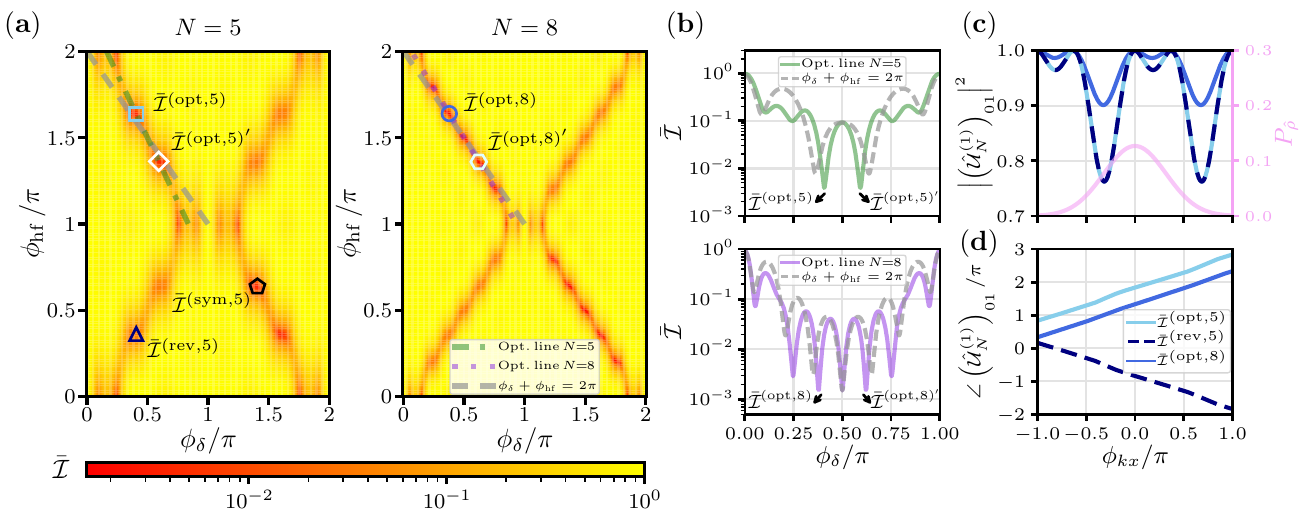}
\caption{(a) Infidelity landscape comparison between protocols with $N=5$ (left) and $N=8$ (right) delta pulses and ion in motional ground state $\ket{f}=\ket{0}$, where the lowest-infidelity points fall in the vicinity of the anti-diagonal line in Eq.~(\ref{eq: anti-diagonal}). $\bar{\mathcal{I}}^{(\text{opt},N)}$ and $\bar{\mathcal{I}}^{(\text{opt},N)'}$ denote the optimal infidelity (that define the optimal line where the local minima are located, which is tilted with respect to the anti-diagonal), $\bar{\mathcal{I}}^{(\text{sym},N)}$ is the corresponding symmetry point, while $\bar{\mathcal{I}}^{(\text{rev},N)}$ denotes the \textquotedblleft reverse-kick\textquotedblright $\,$optimal infidelity that best approximates $\hat{\mathcal{U}}_\text{SDK}(-\phi_{kx})$. (b) 
Cuts comparing local infidelity minima along the anti-diagonal line in Eq.~(\ref{eq: anti-diagonal}) (dashed), and the optimal lines (solid) exhibiting notably better infidelities. (c) Modulus squared of the off-diagonal element  (left axis) $\big(\hat{\mathcal{U}}_N^{(1)}\big)_{01}$ versus $\phi_{kx}$ for the optimal infidelities $\bar{\mathcal{I}}^{(\text{opt},N)}$ and $\bar{\mathcal{I}}^{(\text{rev},N)}$. The right axis represents the position probability distribution of the ground state of motion $\ket{f}=\ket{0}$. (d) Phase of the element $\big(\hat{\mathcal{U}}_N^{(1)}\big)_{01}$ versus $\phi_{kx}$ for $\bar{\mathcal{I}}^{(\text{opt},N)}$ and $\bar{\mathcal{I}}^{(\text{rev},N)}$.}
\label{fig: search}
\end{figure}
We want to find the point(s) that show the smallest infidelity for different number of pulses 
with the ion in a certain motion state. The search space consists of points $\big\{\{\phi_\delta,\,\phi_\text{hf}\}:\phi_\delta,\,\phi_\text{hf}\in[0,\,2\pi)\big\}$, which can be represented in a 2D map, as in Fig.~\ref{fig: search}(a). The infidelity landscape presents some symmetry, a point $\{\phi_\delta,\,\phi_\text{hf}\}$ in the left half of the map has a corresponding point in the right half $\big\{\phi_\delta+\pi,\,\phi_\text{hf}-\pi\pmod{2\pi}\big\}$ with the same infidelity value, simplifying the search for $\phi_\delta\in[0,\,\pi)$ and $\phi_\text{hf}\in[0,\,2\pi)$.

We find that the lower-infidelity points concentrate in bands that lie in the environment of 
\begin{align}\label{eq: diagonal}
    \phi_\delta-\phi_\text{hf} &= 0,
\end{align}
which corresponds to Eq.~(\ref{eq: Mizrahi condition 1}) with equality sign (it simplifies to $\delta=\omega_\text{hf}$, the continuous-wave resonant condition), and
\begin{align}\label{eq: anti-diagonal}
    \phi_\delta+\phi_\text{hf} &= 2\pi,
\end{align}
corresponding to Eq.~(\ref{eq: Mizrahi condition 2}).

We perform an optimization of the infidelity in Eq.~(\ref{eq: I_sc}) for protocols of different number of pulses assuming that the ion is in a given Gaussian motional state, either a coherent state or a thermal state with average number of phonons $\bar{n}=1/\left(\exp\big\{\hbar\omega_\text{t}/(k_BT)\big\}-1\right)$, where $k_B$ is the Boltzmann's constant and $T$ is the temperature. Our optimization strategy uses multiple initial seeds and identifies the optimal solution as the converged solution with the lowest infidelity. Due to symmetry, the seeds are placed in the left half of the map. They include the local infidelity minima along the anti-diagonal line in Eq.~(\ref{eq: anti-diagonal}) (see Fig.~\ref{fig: search}(b)) and the global maxima along the lines $\phi_\text{hf}=\pi$ and $\phi_\text{hf}=2\pi$; see Fig.~\ref{fig: search}(a). We do not incorporate as initial seeds the local maxima along the diagonal line in Eq.~(\ref{eq: diagonal}), as they match the forbidden condition outlined in Eq.~(\ref{eq: Mizrahi condition 1}) and, as will be justified later, correspond to points that approximate the \textquotedblleft reverse-kick\textquotedblright{} SDK operator $\hat{\mathcal{U}}_\text{SDK}(-\phi_{kx})$.

For the discussion of the results we will also refer to two characteristics concerning the off-diagonal element of the pulse train operator in Eq.~(\ref{eq: U_N delta}), $\big(\hat{\mathcal{U}}_N^{(1)}\big)_{01}$, as compared to the corresponding element of the ideal SDK unitary operator in Eq.~(\ref{eq: U_SDK sc}): (i) the squared modulus $\big|\big(\hat{\mathcal{U}}_N^{(1)}\big)_{01}\big|^2$, which ideally should be equal to 1, and (ii) the relationship of the phase $\angle\big(\hat{\mathcal{U}}_N^{(1)}\big)_{01}$ with $\phi_{kx}$, which is linear in the ideal case.

\subsection{Results}

Following the optimization strategy described, we find two optimal solutions, $\big\{\phi_\delta^{(\text{opt},N)},\,\phi_\text{hf}^{(\text{opt},N)}\big\}$ and $\big\{\phi_\delta^{(\text{opt},N)'},\,\phi_\text{hf}^{(\text{opt},N)'}\big\}$, that present the same infidelity value $\bar{\mathcal{I}}^{(\text{opt},N)}$, related by:
\begin{equation}\label{eq: optimal point symmetry}
    \left\{\phi_\delta^{(\text{opt},N)'},\,\phi_\text{hf}^{(\text{opt},N)'}\right\}=\left\{\pi-\phi_\delta^{(\text{opt},N)},\,\pi-\phi_\text{hf}^{(\text{opt},N)}\,\,(\text{mod}\,\,2\pi)\right\},
\end{equation}
and two additional points in the right half of the map that follow from the symmetry relation described:
\begin{equation}
    \left\{\phi_\delta^{(\text{sym},N)},\,\phi_\text{hf}^{(\text{sym},N)}\right\}=\left\{\phi_\delta^{(\text{opt},N)}+\pi,\,\phi_\text{hf}^{(\text{opt},N)}-\pi\,\,(\text{mod}\,\,2\pi)\right\}.
\end{equation}

For a small number of pulses, the \textquotedblleft optimal line\textquotedblright{} containing the local minima is tilted with respect to the ideal $N\rightarrow\infty$ condition in Eq.~(\ref{eq: anti-diagonal})---the anti-diagonal line---and presents visibly better infidelity values; see Fig.~\ref{fig: search}(a-b). Furthermore, we show that there are a limited number of good-infidelity points, contrary to the full span of the anti-diagonal. As the number of pulses increases, the low-infidelity bands become narrower, develop more numerous and deeper local minima, and align more closely with Eqs.~(\ref{eq: diagonal}) and (\ref{eq: anti-diagonal}). This behavior is consistent with Refs.~\cite{mizrahi2013,mizrahi2014}, which predict vanishing infidelity in the anti-diagonal region in the limit of an infinite number of pulses.

\begin{figure}[t!]
\centering
\includegraphics[width=0.5\textwidth]{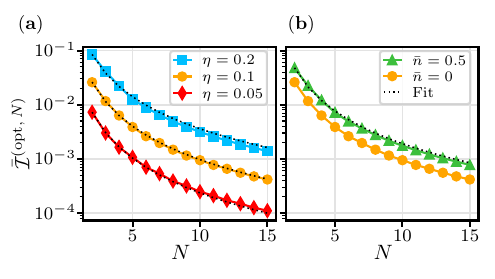}
\caption{Delta-pulse optimal state-averaged infidelity versus pulse number in the protocol. (a) Results for ion in motional ground state $\ket{f}=\ket{0}$ for different values of the Lamb--Dicke parameter. The fit parameters $\{K,\,r\}$ of the power-law decay $\bar{\mathcal{I}}^{(\text{opt},N)}=KN^{-r}$ (dotted lines) are $\{0.33,\,1.96\}$, $\{0.11,\,2.05\}$ and $\{0.03,\,2.13\}$ for $\eta=0.2,\,0.1,\,0.05$, respectively. (b) Comparison of the optimal infidelity for the ground state $(\bar{n}=0)$ and a thermal state with $\bar{n}=0.5$, both for $\eta=0.1$. The fit parameters for $\bar{n}=0.5$ are $\{0.19,\,2.00\}$.}
\label{fig: delta pulse opt. F}
\end{figure}
We find that the optimal infidelity $\bar{\mathcal{I}}^{(\text{opt},N)}$ follows a power-law decay with increasing pulse number: $\bar{\mathcal{I}}^{(\text{opt},N)}=KN^{-r}$, with $K,\,r\in\mathbb{R}$, as shown in Fig.~\ref{fig: delta pulse opt. F}. $\bar{\mathcal{I}}^{(\text{opt},N)}$ becomes greater with increasing width of the ionic motional wavepacket, caused either by a larger Lamb--Dicke parameter or higher temperature. This is due to the SDK protocol not having the same quality at every position throughout the trap. Therefore, the state-averaged infidelity will be smaller if the ion is more localized in positions where the overlap $\bra{Q}\hat{\mathcal{U}}_\text{SDK}^\dagger\hat{\mathcal{U}}_N^{(1)}\ket{Q}$ is maximal.

This position dependence is manifested in the value $\big|\big(\hat{\mathcal{U}}_N^{(1)}\big)_{01}\big|^2$ for the optimal infidelities $\bar{\mathcal{I}}^{(\mathrm{opt},N)}$. It presents a maximum coinciding with the ideal one at the position where it is most likely to find the ion, and a decrease for farther values of $x$; see Fig.~\ref{fig: search}(c). The higher optimal infidelity seen for the protocol with fewer pulses is evidenced by the greater deviation of $\big|\big(\hat{\mathcal{U}}_N^{(1)}\big)_{01}\big|^2$ from 1 as we move away from $x=0$.

The phase of $\big(\hat{\mathcal{U}}_N^{(1)}\big)_{01}$ shows an approximate linear relationship with $\phi_{kx}$, as depicted in Fig.~\ref{fig: search}(d). A root-mean-square measure of this linearity reveals a better adjustment for the protocol with a higher pulse number. The y-intercept unveils the value of $\xi$ in the definition of $\hat{\mathcal{U}}_\text{SDK}$ in Eq.~(\ref{eq: U_SDK sc}). Points lying around the diagonal line in Eq.~(\ref{eq: diagonal}) show a reverted kick direction, as evidenced by the negative slope. The pair $\big\{\phi_\delta^{(\text{rev},N)},\,\phi_\text{hf}^{(\text{rev},N)}\big\}$, with $\phi_\delta^{(\text{rev},N)}=\phi_\delta^{(\text{opt},N)}$ and $\phi_\text{hf}^{(\text{rev},N)}=2\pi-\phi_\text{hf}^{(\text{opt},N)}$, constitutes the optimal point if we target $\hat{\mathcal{U}}_\text{SDK}(-\phi_{kx})$, which corresponds to a SDK operator with the kick direction exchanged for each spin state. This explains the local infidelity maximum at $\{\pi,\,\pi\}$, consistent with Eqs.~(\ref{eq: Mizrahi condition 1}) and (\ref{eq: Mizrahi condition 2}), which must not be fulfilled simultaneously. The infinite-pulse number exact conditions for having $\hat{\mathcal{U}}_\text{SDK}(-\phi_{kx})$ result from swapping the \textquotedblleft$\ne$\textquotedblright{} and \textquotedblleft$=$\textquotedblright{} signs in Eqs.~(\ref{eq: Mizrahi condition 1}) and (\ref{eq: Mizrahi condition 2}), respectively~\cite{mizrahi2013,mizrahi2014}.

For a given atomic isotope of mass $M$ excited by light with wavenumber $k$ the Lamb--Dicke parameter $\eta$ can be reduced by increasing the trap frequency. $\eta$ is an important parameter, since faster two-qubit gate schemes that rely on sequences of SDKs require either more SDKs or smaller trap frequencies~\cite{wong-campos2017}. In addition, there exists a trade-off between kick strength $\eta u$ (required for enough geometric phase accumulation in two-qubit gate schemes~\cite{mizrahi2014}) and fidelity.

These results illustrate how the semiclassical approach provides more insight into the dependence of infidelity on the Lamb--Dicke parameter and temperature.

\section{Finite-width pulse protocols}\label{sec:finite pulses}

We now consider finite pulse train protocols with pulses of finite width in the absence of ion motion, to quantify the impact of neglecting the pulse temporal profile on the infidelity. Trapped-ion hyperfine qubits have typical frequencies $\omega_\text{hf}\sim\!1\text{--}20\,\mathrm{GHz}$~\cite{NAP2019TrappedIon} so, if excited by a mode-locked picosecond laser source (with pulse duration in the range $\tau\sim1\text{--}20\,\mathrm{ps}$), the hyperfine period and the pulse duration timescales are not well separated, and this will prove to be the dominant source of error. This differs from the ideal dynamics of the few instantaneous pulses considered in the previous section.

\subsection{Formulation}

We still neglect the trap potential term and now lift the instantaneous-pulse approximation, so we consider the pulse pair Hamiltonian in Eq.~(\ref{eq: H_p parameter}) with the more realistic pulse envelope modeled as a Gaussian profile
\begin{equation}\label{eq: Gaussian envelope}
    \mathcal{A}(t)=\frac{\theta}{\sqrt{2\pi\sigma^2}}e^{-\frac{t^2}{2\sigma^2}},
\end{equation}
where $\sigma^2$ is the variance, and the pulse width $\tau=\sigma\sqrt{8\log2}$ is taken as the full-width at half-maximum of the Gaussian.

The temporal dependence of the pulse envelope in Eq.~(\ref{eq: Gaussian envelope}) does not allow for a closed-form analytical solution of the evolution operator for a single finite-width pulse pair, as was the case for delta pulses. Therefore, we approximate the propagator numerically using Trotterization. For the considered extension of the pulse $t_\text{ext}=m\sigma$, where $m$ is the number of standard deviations spanning the truncated Gaussian, the pulse pair operator is computed as
\begin{equation}
    \hat{\mathcal{U}}^{(2)}(t,\,t_n,\,\phi_{kx})=\prod_{s=0}^{S-1}\exp\left\{-i\left[\hat{H}_\text{hf} + \hat{H}_\text{pulse}\left(t_s+dt/2,\,t_n,\,\phi_{kx}\right)\right]dt\right\},
\end{equation}
where $t_\text{s}=t_n-t_\text{ext}/2+sdt$, $S$ is the number of Trotter steps and $dt=t_\text{ext}/S$ is the differential time step.
The pulse train operator thus reads
\begin{equation}\label{eq: U_N finite}
    \hat{\mathcal{U}}_N^{(2)}(t,\,\phi_{kx})=\prod_{n=0}^{N-2}\left[\hat{\mathcal{U}}^{(2)}(t,\,t_n,\,\phi_{kx})\hat{\mathcal{U}}_\text{FE}^\text{(hf)}(t_\text{rep}-t_\text{ext})\right]\hat{\mathcal{U}}^{(2)}(t,\,t_0,\,\phi_{kx}).
\end{equation}
We take the same expression for the target SDK operator given in Eq.~(\ref{eq: U_SDK sc}) and the fidelity in Eq.~(\ref{eq: F sc}) with the corresponding operator $\hat{\mathcal{U}}_N^{(2)}$.

\subsection{Methodology}

At the level of a single pulse pair, the non-commuting qubit $\big(\hat{H}_\text{hf}\big)$ and pulse pair $\big(\hat{H}_\text{pulse}(t,\,t_n,\,\phi_{kx})\big)$ terms introduce an error in the interaction dynamics relative to the ideal delta-pulse scenario. We use time-dependent perturbation theory to compute an error bound due to the finite duration of each pulse, where the slow $\hat{H}_\text{hf}$ term in Eq.~\eqref{eq: H} is treated perturbatively with respect to $\hat{H}_\text{pulse}(t,\,t_n,\,\phi_{kx})$. To this end, we express the pulse pair evolution operator as
\begin{equation}\label{eq:U0U1 error}
    \tilde{\hat{\mathcal{U}}}^{(2)}(t,\,t_n,\,\phi_{kx})=\hat{\mathcal{U}}_\text{pulse}(t,\,t_n,\,\phi_{kx})\hat{\mathcal{U}}_\text{hf}(t,\,t_n,\,\phi_{kx}),
\end{equation}
where
\begin{align}
    \hat{\mathcal{U}}_\text{pulse}(t,\,t_n,\,\phi_{kx})&=\exp\left\{-i\int_{t_n-t_\text{ext}/2}^t\hat{H}_\text{pulse}(t,\,t_n,\,\phi_{kx})dt'\right\},\\
    \frac{d}{dt}\hat{\mathcal{U}}_\text{hf}(t,\,t_n,\,\phi_{kx})&=-i\hat{H}_\text{hf}^{(\text{I})}(t,\,t_n,\,\phi_{kx})\,\hat{\mathcal{U}}_\text{hf}(t,\,t_n,\,\phi_{kx}),\label{eq:dU_hf/dt}\\
    \hat{H}_\text{hf}^{(\text{I})}(t,\,t_n,\,\phi_{kx})&=\hat{\mathcal{U}}_\text{pulse}^\dagger(t,\,t_n,\,\phi_{kx})\,\hat{H}_\text{hf}\,\hat{\mathcal{U}}_\text{pulse}(t,\,t_n,\,\phi_{kx}).
\end{align}
If $\hat{H}_\text{pulse}(t,\,t_n,\,\phi_{kx})$ and $\hat{H}_\text{hf}$ commuted, we would have $\hat{\mathcal{U}}_\text{hf}(t,\,t_n)=\hat{\mathcal{U}}_\text{FE}^\text{(hf)}\left(t-\left(t_n-t_\text{ext}/2\right)\right)$; however, that is not the case, so we approximate $\hat{\mathcal{U}}_\text{hf}(t,\,t_n,\,\phi_{kx})\approx\tilde{\hat{\mathcal{U}}}_\text{hf}(t,\,t_n)$ using the Magnus expansion. The single-pulse pair error is thus defined as
\begin{equation}\label{eq:error finite}
    \tilde{\epsilon}_{\scriptscriptstyle \Omega}=\left\|\hat{\mathcal{U}}_\text{FE}^\text{(hf)}(\tau)-\tilde{\hat{\mathcal{U}}}_\text{hf}(t_n+\tau/2,\,t_n)\right\|,
\end{equation}
where $\|\cdot\|$ denotes matrix norm. It can be approximated as (see Appendix~\ref{Appendix A})
\begin{equation}\label{eq: error finite approx}
    \tilde{\epsilon}_{\scriptscriptstyle \Omega}(\tau,\,N)\approx\frac{\omega_\text{hf}\tau}{2}\sqrt{1-\text{sinc}\left(\frac{\gamma}{N}\right)\left[2\cos\left(\frac{\beta}{N}\right)-\text{sinc}\left(\frac{\gamma}{N}\right)\right]},
\end{equation}
where $\beta=\pi\,\text{erf}\left(-\sqrt{\log2}\right)$, $\gamma=2\sqrt{\pi\log2}$, and $\text{sinc}\left(a\right)=\sin(a)/a$.
The error propagates for each pulse pair, so the total error bound is
\begin{equation}
    1-\left[1-\tilde{\epsilon}_{\scriptscriptstyle \Omega}(\tau,\,N)\right]^N\approx1-\left[1-N\tilde{\epsilon}_{\scriptscriptstyle \Omega}(\tau,\,N)\right]=N\tilde{\epsilon}_{\scriptscriptstyle \Omega}(\tau,\,N),
\end{equation}
where we have approximated $\left[1-\tilde{\epsilon}_{\scriptscriptstyle \Omega}(\tau,\,N)\right]^N$ to first order in $\tilde{\epsilon}_{\scriptscriptstyle \Omega}(\tau,\,N)$. The dependence with the number of pulses for large $N$ vanishes to leading order
\begin{equation}
    N\tilde{\epsilon}_{\scriptscriptstyle \Omega}(\tau,\,N)\approx\frac{\omega_\text{hf}\tau}{2}|\beta|+\mathcal{O}\left(N^{-2}\right).
\end{equation}
We compare this error bound with numerical simulations of the pulse train operator in Eq.~(\ref{eq: U_N finite}) evaluated at the delta-pulse optimal points.

\subsection{Results}

\begin{figure}[t!]
\centering
\includegraphics[width=0.5\textwidth]{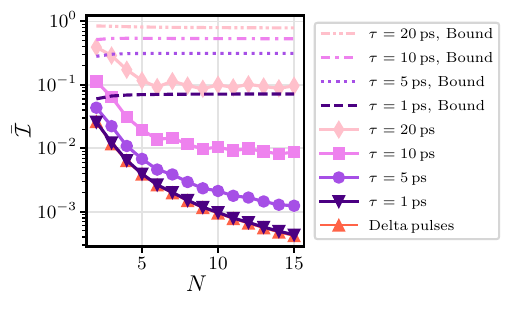}
\caption{Infidelity versus pulse number evaluated at $\big\{\phi_\delta^{(\text{opt},N)},\,\phi_\text{hf}^{(\text{opt},N)}\big\}$ for different pulse widths (the points $\big\{\phi_\delta^{(\text{opt},N)'},\,\phi_\text{hf}^{(\text{opt},N)'}\big\}$ exhibit larger deterioration, not shown) and error bound $N\tilde{\epsilon}_{\scriptscriptstyle \Omega}$. We have considered a $^{133}\text{Ba}^+$ ion $(\omega_\text{hf}\approx2\pi\times10\,\mathrm{GHz})$ in the ground state of motion $\ket{f}=\ket{0}$ with $\eta=0.1$.}
\label{fig: finite pulse error}
\end{figure}
We simulate the protocol with finite-width pulses introducing the ion species $^{133}\text{Ba}^+$ $(\omega_\text{hf}\approx2\pi\times10\,\mathrm{GHz})$ without loss of generality.

We observe that the infidelity of the protocol with pulses of finite width is degraded with increasing pulse duration with respect to the value obtained for delta pulses, except for very short pulses $\tau\sim1\text{--}5\,\mathrm{ps}$;
see Fig.~\ref{fig: finite pulse error}. In this case, the product of the hyperfine frequency and the pulse width yields $\tau\omega_\text{hf}/(2\pi)=\!1\text{--}5\times10^{-2}$, which serves as an indicator to evaluate the validity of the delta-pulse approximation. This is a consequence of the different orientations in Pauli space of $\hat{H}_\text{hf}$ and $\hat{H}_\text{pulse}(t,\,t_n,\,\phi_{kx})$, which cause the pulse pair to generate a rotation of the qubit about an axis distinct from the pure $\hat{\sigma}_x$-rotation present in the delta-pulse limit. This demonstrates that the optimal points found assuming instantaneous pulses are not necessarily optimal for finite-width pulse protocols and are thus displaced.

The single-pulse pair error accumulates as more pulses are added so, for a given pulse width, the error relative to the delta-pulse infidelity increases with pulse number. We find a $100\%$ relative error for protocols with $7\text{--}9$ pulses and pulse duration as low as $5\,\mathrm{ps}$. The infidelity suffers an order of magnitude rise (with respect to the delta-pulse value) with $10$ pulses $10\,\mathrm{ps}$-long, whereas it becomes two orders of magnitude greater with $20\,\mathrm{ps}$ pulses. Even though the analytic bound is pessimistic, it correctly predicts the infidelity saturation observed for sufficiently large pulse number. In contrast to the delta-pulse case, where the infidelity continues to improve with increasing pulse number, there is no such improvement for higher-width pulses, as the gain is counteracted by the pulse width error.

\section{Ion motion}\label{sec:ion motion}

\begin{figure}
\centering
\includegraphics[width=0.5\textwidth]{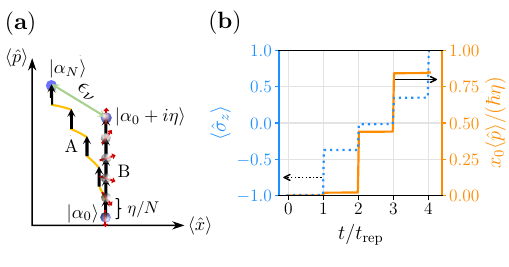}
\caption{(a)  Phase space picture of a $5$-pulse SDK with (path A) and without (path B) taking into account the secular motion of the ion, assuming equal-strength momentum kicks  $p=\eta/N$ (in units of $\hbar/x_0$) depicted as black arrows. The error $\epsilon_\nu$ is regarded as the distance between the final states $\ket{\alpha_N}$ and $\ket{\alpha_0+i\eta}$. The red arrows represent the spin state, which rotates with every momentum displacement (only shown in path B for visibility). (b) SDK simulation including ion motion for $N=5$ pulses of duration $5\,\mathrm{ps}$ at the delta-pulse optimal point with $t_\text{rep}\approx0.28\,\mathrm{ns}$ in a trap of frequency $\omega_\text{t}=2\pi\times2\,\mathrm{MHz}$. The expectation values $\langle\hat{\sigma}_z\rangle$ (left axis, dotted) and $\langle\hat{p}\rangle$ (right axis, solid) are depicted throughout the SDK time. In contrast to the simple process illustrated in panel (a), the momentum kick provided by each pulse has a different value; in this particular case, only the third and fourth pulses mainly contribute to the momentum transfer.}

\label{fig: ion motion}
\end{figure}

Having considered in the previous section the real characteristics of picosecond pulses, we now lift the zero-motion approximation and we simulate the complete model. In the timescale of the SDK time around a few nanoseconds long, ion motion is expected to be small for typical trap frequencies in the megahertz range; nonetheless we characterize the influence of ion secular motion on the infidelity, both analytically and numerically, finding a negligible error compared with that due to the finite pulse duration.

\subsection{Formulation}

For convenience, we restate the complete Hamiltonian in Eq.~(\ref{eq: H}) $(\varphi_0=0)$
\begin{equation}\label{eq: H2}
    \hat{H}=\omega_\text{t}\hat{a}^\dagger\hat{a} + \frac{\omega_\text{hf}}{2}\hat{\sigma}_z + \mathcal{W}(t)\sin\left(\eta\big(\hat{a}^\dagger+\hat{a}\big) + \delta t\right)\hat{\sigma}_x.
\end{equation}
The ion is allowed to move in a harmonic oscillator potential and the pulse interaction term, given in Eq.~(\ref{eq: H_p operator}), has the position as a quantum operator.

In this section, we assume that the ion is initially in a spin--coherent state $\ket{Q,\alpha_0}$, so the final evolved state is given by $\hat{\mathcal{U}}_N^{(3)}\ket{Q,\alpha_0}$ (details on the state propagation given in Appendix~\ref{sec:Appendix B}).
The infidelity in the combined spin--motion Hilbert space is defined as
\begin{equation}\label{eq:F spin--motion}
    \mathcal{I}=1-\left|\bra{Q,\alpha_0}\hat{U}_\text{SDK}^\dagger\hat{\mathcal{U}}_N^{(3)}\ket{Q,\alpha_0}\right|^2.
\end{equation}

\subsection{Methodology}

The trap potential $\big(\hat{H}_\text{t}\big)$ and pulse pair $\big(\hat{H}_\text{pulse}(t,\,t_n)\big)$ hamiltonians do not commute; however, we have $\tau\int\mathcal{A}(t-t_n)dt\gg\omega_\text{t}\tau$ by $5\text{--}6$ orders of magnitude. Therefore, for the theoretical error estimation we assume the delta-pulse scenario, in which the ion gets $N$ instantaneous momentum kicks separated with harmonic free evolution for a time $t_\text{rep}$. The error is thus measured as
\begin{equation}\label{eq:error motion}
    \tilde{\epsilon}_\nu=1-\left|\langle \alpha_0+i\eta|\alpha_N\rangle\right|^2,
\end{equation}
where $\alpha_0$ is the amplitude of the initial coherent state and $\ket{\alpha_N}$ is the idealized final coherent state after $N$ kicks of momentum $p=\eta/N$ (in units of $\hbar/x_0$) with motion between pulses; see Fig.~\ref{fig: ion motion}(a). The state $\ket{\alpha_0+i\eta}$ corresponds to the target motional state as outlined in Eq.~(\ref{eq: target state}). The trap potential term $\hat{H}_\text{t}$ does not affect the spin state, so in Eq.~(\ref{eq:error motion}) we considered the same final spin state for the evolved states with and without motion. Equation~(\ref{eq:error motion}) can be approximated to (see Appendix~\ref{sec:Appendix C})
\begin{equation}\label{eq: error motion approx}
    \tilde{\epsilon}_\nu\approx\left(\frac{\eta\omega_\text{t}t_\text{rep}(N-1)}{2}\right)^2.
\end{equation}
The actual infidelity error caused by the motion is measured as
\begin{equation}
    \epsilon_\nu=\mathcal{I}^{(\nu)}-\mathcal{I}^{(\Omega)},
\end{equation}
where $\mathcal{I}^{(\nu)}$ and  $\mathcal{I}^{(\Omega)}$ are the infidelities with and without ion motion, respectively, at the delta-pulse optimal points.

Contrary to the theoretical error calculation where we assumed instantaneous pulses, we perform simulations considering finite-duration pulses with and without the trap potential term $\hat{H}_\text{t}$. We extract the repetition time $t_\text{rep}$ and the Raman frequency difference $\delta$ from the delta-pulse optimal points and evaluate the infidelity in Eq.~(\ref{eq:F spin--motion}). 

\begin{figure}[t]
\centering
\includegraphics[width=0.5\textwidth]{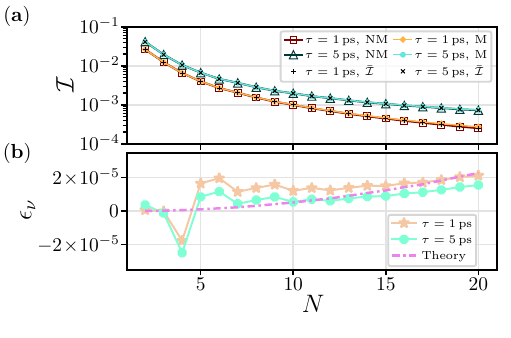}
\caption{SDK simulation for the ion initially in the state $\ket{\Psi_0}=\ket{1}\ket{0}$ with $\big(\text{M, }\mathcal{I}^{(\nu)}\big)$ and without $\big(\text{NM, }\mathcal{I}^{(\Omega)}\big)$ ion motion at the delta-pulse optimal points for different pulse durations. The parameter values are $t_\text{rep}\approx0.25\text{--}0.28\,\mathrm{ns}$, $\omega_\text{hf}\approx2\pi\times10\,\mathrm{GHz}$ and $\omega_\text{t}=2\pi\times2\,\mathrm{MHz}$. (a) Infidelity versus pulse number compared to the semiclassical values $\bar{\mathcal{I}}$, where position is treated as a parameter. (b) Difference between infidelities with and without ion motion, and theoretical error estimation $\tilde{\epsilon}_\nu$ in Eq.~(\ref{eq: error motion approx}).}
\label{fig: ion motion error}
\end{figure}

\subsection{Results}

In our simulations we consider a $^{133}\text{Ba}^+$ ion ($\omega_\text{hf}\approx2\pi\times10\,\mathrm{GHz}$) in a $\omega_\text{t}=2\pi\times2\,\mathrm{MHz}$ trap excited by light with center wavelength $\lambda=532\,\mathrm{nm}$ (each beam is shifted in frequency by $\pm\delta/2$), so that the Lamb--Dicke parameter is $\eta\approx0.1$.

For $t_\text{rep}$ of fractions of $\mathrm{ns}$ (so that $T_\text{SDK}\sim\mathrm{ns}$, as in Refs. ~\cite{mizrahi2013,johnson2017,wong-campos2017}) we obtain minimal deterioration of the infidelity when the motion of the ion is included; see Fig.~\ref{fig: ion motion error}. The error is of the order of $10^{-5}$ for $T_\text{SDK}\omega_\text{t}/(2\pi)=5\text{--}5.6\times10^{-4}N$, orders of magnitude smaller than the dominating error due to the finite pulse duration. This is due to the ion falling in trap zones where the protocol deviates from the ideal operation. Moreover, we find excellent agreement with the semiclassical infidelity defined in Eq.~(\ref{eq: I_sc}), which proves to be an accurate description when the motion of the ion does not play a role in the dynamics.

We identify two regions with different error trends in Fig.~\ref{fig: ion motion error}(b). For $N\gtrsim7$ we see a clear increasing tendency as more pulses are added to the realization of the SDK---since the larger the protocol time $T_\text{SDK}$ the more the ion moves---which is closely followed by the analytical error in Eq.~(\ref{eq: error motion approx}). This matching is more apparent for $\tau=1\,\mathrm{ps}$, as the pulses behave effectively as delta pulses for the considered $\omega_\text{hf}\approx2\pi\times10\,\mathrm{GHz}$. We observe a smaller error $\epsilon_\nu$ for broader pulses, which is corroborated numerically due to a smaller ion displacement. For $N\lesssim7$ the error does not have a predictable behavior, although it remains within the same order of magnitude anticipated by the theoretical error in Eq.~(\ref{eq: error motion approx}). Due to the temporal dependence of the sine term in the Rabi frequency (see Eq.~(\ref{eq: H2})), the momentum provided by each pulse is different, as represented in Fig.~\ref{fig: ion motion}(b), and is more evenly distributed among pulses for protocols with larger $N$, which differs from the simple scheme depicted in Fig.~\ref{fig: ion motion}(a). As a consequence, the motion of the ion is almost always larger than the theoretical prediction (it has the opposite behavior for $\ket{\Psi_0}=\ket{0}\ket{0}$, not represented). 

\section{Conclusions and outlook}\label{sec:conclusions}

We accurately find the optimal points $\{\phi_\delta,\,\phi_\text{hf}\}$ (related to the frequency difference between the Raman beams, the repetition time between the pulses and the hyperfine qubit frequency) by optimizing SDK infidelity on a stationary ion with the exact, few-delta-pulse propagator (with equispaced, equal-amplitude pulses), which differ from the large pulse number limit conditions~\cite{mizrahi2013,mizrahi2014}.

The optimal infidelity follows a power-law decay with the number of pulses, which rapidly goes below $10^{-3}$ for $\sim10$ pulses (with $\eta=0.1$ and the ion in the ground state of motion). We find that the optimal infidelity increases as the motional wavepacket of the ion becomes wider either by a larger Lamb--Dicke parameter or higher temperature, which displaces the optimal points, and is easily inferred by our semiclassical approach.

We show that SDK infidelity in few-pulse protocols is dominated by finite-pulse duration and not ion motion.
The instantaneous-pulse approximation taken in the optimization is only reliable for very short pulses, when $\tau\omega_\text{hf}/(2\pi)\sim\!1\text{--}5\times10^{-2}$, which translates to pulse durations $\tau\sim1\text{--}5\,\mathrm{ps}$ for hyperfine frequencies $\omega_\text{hf}\approx2\pi\times10\,\mathrm{GHz}$. The infidelity error relative to the ideal delta-pulse case deteriorates by one (two) orders of magnitude for $10$ pulses of duration $10\,\mathrm{ps}$ ($20\,\mathrm{ps}$), corresponding to conditions similar to performed experiments~\cite{mizrahi2013,johnson2017}. We confirm that ion secular motion introduces a small error of the order of $10^{-5}$ for typical SDK times $T_\text{SDK}\!\sim\!\mathrm{ns}$ where the trap frequency is $\omega_\text{t}\!\sim\!\mathrm{MHz}$ ($T_\text{SDK}\omega_\text{t}/(2\pi)=5\text{--}5.6\times10^{-4}N$).

Future work could explore ways to increase the fidelity of SDKs. The infidelity deterioration due to the finite pulse duration can be approximately compensated by an increase in the pulse area at the expense of higher laser power~\cite{mizrahi2014}. As an improved alternative, due to the critical impact of pulse width, a new optimization of the parameters $\{\phi_\delta,\,\phi_\text{hf}\}$ could be performed accounting for the finite pulse duration. Furthermore, higher fidelities may be achieved if the optimization space is enlarged by having unequal pulse amplitudes and/or arrival times~\cite{HLiu2026,Torrontegui20}, limited by the error bound due to the ion motion that we find here.

This work establishes the relevant aspects to consider when optimizing SDKs with short pulses and provides both theoretical error scalings and actual simulated values of infidelity due to the instantaneous-pulse and frozen-ion approximations. These results determines the optimal working conditions, establishing the path towards the generation of fast two-qubit entangling gates with pulsed sources.

\appendix

\suppdata{{\color{white} .}}

\section{Finite-width pulse error estimation}\label{Appendix A}

We want to approximate $\hat{\mathcal{U}}_\text{hf}(t,\,t_n,\,\phi_{kx})$ in Eq.~(\ref{eq:dU_hf/dt}) using the Magnus expansion to first order
\begin{equation}
    \tilde{\hat{\mathcal{U}}}_\text{hf}(t,\,t_n,\,\phi_{kx})=e^{\Omega_1(t,\,t_n,\,\phi_{kx})},
\end{equation}
where
\begin{equation}\label{eq: Magnus1}
    \Omega_1(t,\,t_n,\,\phi_{kx})=-i\int_{t_n-\tau/2}^{t} \hat{H}_\text{hf}^{(\text{I})}\left(t'',\,t_n,\,\phi_{kx}\right)dt''.
\end{equation}
The Hamiltonian $\hat{H}_\text{hf}$ in the interaction picture with respect to $\hat{H}_\text{pulse}(t'',\,t_n,\,\phi_{kx})$ can be expressed as
\begin{align}
    \hat{H}_\text{hf}^{(\text{I})}(t,\,t_n,\,\phi_{kx})&=e^{i\zeta(t,\,t_n,\,\phi_{kx})\hat{\sigma}_x}\,\hat{H}_\text{hf}\,e^{-i\zeta(t,\,t_n,\,\phi_{kx})\hat{\sigma}_x}\nonumber\\
    &=\frac{\omega_\text{hf}}{2}\left(\cos\left[2\zeta\left(t,\,t_n,\,\phi_{kx}\right)\right]\hat{\sigma}_z+\sin\left[2\zeta\left(t,\,t_n,\,\phi_{kx}\right)\right]\hat{\sigma}_y\right),
\end{align}
where
\begin{equation}
    \zeta(t,\,t_n,\,\phi_{kx})=\int_{t_n-\tau/2}^t\hat{H}_\text{pulse}\left(t',\,t_n,\,\phi_{kx}\right)dt'.
\end{equation}
A sufficient condition for convergence of the series is
\begin{equation}
    \pi>\int_{t_n-\tau/2}^{t_n+\tau/2}\left\|\hat{H}_\text{hf}^{(\text{I})}(t,\,t_n,\,\phi_{kx})\right\|=\frac{\omega_\text{hf}}{2}\int_{t_n-\tau/2}^{t_n+\tau/2}dt_1=\frac{\omega_\text{hf}\tau}{2},
\end{equation}
which is fulfilled for $\tau\lesssim100\,\mathrm{ps}$ if $\omega_\text{hf}\approx2\pi\times10\,\mathrm{GHz}$. To compute the matrix norm we used $\|A\|=|c_1|+\sqrt{|c_x|^2+|c_y|^2+|c_z|^2}$, for any $2\times2$ real matrix $A=c_1\mathbb{I}+c_x\sigma_x+c_y\sigma_y+c_z\sigma_z$ in the basis $\{\mathbb{I},\,\sigma_x,\,\sigma_y,\,\sigma_z\}$. Considering the Gaussian pulse envelope in Eq.~(\ref{eq: Gaussian envelope}) we have
\begin{align}\label{eq: xi approx}
    \zeta(t,\,t_n)&\approx\frac{\theta\sin\big(\phi_{kx}+\delta t_n\big)}{\sigma\sqrt{2\pi}}\int_{t_n-\tau/2}^{t}\exp\left\{\frac{-\left(t'-t_n\right)^2}{2\sigma^2}\right\}dt'\nonumber\\
    &\approx\frac{\theta'}{2}\left[\frac{1}{\sigma}\sqrt{\frac{2}{\pi}}(t-t_n)-\frac{\beta}{\pi}\right]\le\frac{\theta}{2}\left[\frac{1}{\sigma}\sqrt{\frac{2}{\pi}}(t-t_n)-\frac{\beta}{\pi}\right].
\end{align}
In the first line of Eq.~(\ref{eq: xi approx}), we assumed $t-\left(t_n-\tau/2\right)\ll2\pi/\delta$, so that the sine term can be treated as a constant. In the second line, we first Taylor-expanded $\text{erf}\left[(t-t_n)/\left(\sigma\sqrt{2}\right)\right]$ to first order about $t=t_n$, and we then bounded $\theta'=\theta\sin\big(\phi_{kx}+\delta t_n\big)\le\theta$ (thereby removing the cumbersome dependence on $\phi_{kx}$). Therefore, we can compute the first order term in Eq.~(\ref{eq: Magnus1}) as
\begin{equation}
    \Omega_1(\tau,\,N)=-i\frac{\omega_\text{hf}\tau}{2}\text{sinc}\left(\frac{\gamma}{N}\right)\left[\cos\left(\frac{\beta}{N}\right)\hat{\sigma}_z-\sin\left(\frac{\beta}{N}\right)\hat{\sigma}_y\right].
\end{equation}
Using the identity $e^{-i\,\mathbf{R}\cdot\boldsymbol{\sigma}}=\cos(R)\mathbb{I}-i\sin(R)\mathbf{R}\cdot\boldsymbol{\sigma}/R$, where $\mathbf{R}=(R_x,\,R_y,\,R_z)$, $\boldsymbol{\sigma}$ is the vector of Pauli matrices and $R=|\mathbf{R}|$, we expand the operators in Eq.~(\ref{eq:error finite}) to first order in $\omega_\text{hf}\tau$. Then, we  compute the matrix norm to arrive to Eq.~(\ref{eq: error finite approx}).

\section{Complete model state propagation}\label{sec:Appendix B}

Instead of finding the approximate pulse train operator $\hat{\mathcal{U}}_N^{(3)}$, we propagate the state using Trotterization for numerical stability. For the extension of the pulses considered $t\in[t_n-t_\text{ext}/2,\,t_n+t_\text{ext}/2]$, the state at each time step during the $n$-th pulse pair is
\begin{equation}
    |\Psi\rangle_{t_{s+1}}^{(t_n)}=\exp\left\{-i\left[\hat{H}_\text{t}+\hat{H}_\text{hf}+\hat{H}_\text{pulse}\left(t_s+dt/2,\,t_n,\,\phi_{kx}\right)\right]dt\right\}\ket{\Psi}_{t_s}^{(t_n)},
\end{equation}
where $t_\text{s}=t_n-t_\text{ext}/2+sdt$, with $s=0,\,...,\,S$, and $\ket{\Psi}_{t_0-t_\text{ext}/2}^{(t_0)}=\ket{Q}\ket{\alpha_0}$. The wavefunction at the end of the pulse dynamics is therefore $\ket{\Psi}_{t_n+t_\text{ext}/2}^{(t_n)}$.
Next, we concatenate both qubit and trap free evolution operators for time $\mathcal{T}=t_\text{rep}-t_\text{ext}$ before the beginning of the next pulse at $t_s=t_{n+1}-t_\text{ext}/2$ as
\begin{equation}
    \ket{\Psi}_{t_{n+1}-t_\text{ext}/2}^{(t_{n+1})}=\hat{\mathcal{U}}_\text{FE}^\text{(hf)}(t_\text{rep}-t_\text{ext})\hat{\mathcal{U}}_\text{FE}^\text{(t)}(t_\text{rep}-t_\text{ext})\ket{\Psi}_{t_n+t_\text{ext}/2}^{(t_n)},
\end{equation}
where we define
\begin{equation}\label{eq: U_FE motion}
    \hat{\mathcal{U}}_\text{FE}^\text{(t)}(\mathcal{T})=e^{-i\omega_\text{t}\mathcal{T}\hat{a}^\dagger\hat{a}}.
\end{equation}

\section{Ion motion error estimation}\label{sec:Appendix C}

To estimate the error in the design of a SDK when the motion of the ion is considered in the evolution, we assume an ideal delta-pulse train which produces a perfect spin flip and every pulse kicks the ion with momentum $p=\eta/N$, so the exact evolution operator $\hat{\mathcal{U}}_\text{N}^{(3)}$ can be approximated by
\begin{equation}
    \tilde{\hat{\mathcal{U}}}_\text{N}^{(3)}=\hat{\sigma}_x\,\otimes\,\left(\prod_{n=0}^{N-2}\left[\hat{\mathcal{D}}(ip)\hat{\mathcal{U}}_\text{FE}^\text{(t)}(t_\text{rep})\right]\hat{\mathcal{D}}(ip)\right).
\end{equation}
Using
\begin{equation}
    \hat{\mathcal{D}}(ip)\ket{\alpha}=\ket{\alpha+ip},
\end{equation}
\begin{equation}
    \hat{\mathcal{U}}_\text{FE}^\text{(t)}(t_\text{rep})\ket{\alpha}=e^{-i\omega_\text{t}t_\text{rep}/2}\ket{\alpha e^{-i\omega_\text{t}t_\text{rep}}},
\end{equation}
\begin{equation}
    \nonumber
\end{equation}
we find recursively $\tilde{\hat{\mathcal{U}}}_\text{N}^{(3)}\ket{Q}\ket{\alpha_0}=e^{-iN\omega_\text{t}t_\text{rep}/2}\ket{Q\oplus1}\ket{\alpha_N}$, with
\begin{align}
    \alpha_N&=\alpha_0e^{-iN\omega_\text{t}t_\text{rep}}+ip\sum_{n=0}^{N-1}e^{-i\omega_\text{t}t_n}\nonumber\\
    &=\alpha_0e^{-iN\omega_\text{t}t_\text{rep}}+i\frac{\eta}{N}e^{-i\omega_\text{t}t_\text{rep}(N-1)/2}\frac{\sin\left(N\omega_\text{t}t_\text{rep}/2\right)}{\sin\left(\omega_\text{t}t_\text{rep}/2\right)}.
\end{align}
The error in Eq.~(\ref{eq:error motion}) can be written as
\begin{equation}
    \tilde{\epsilon}_\nu=1-\left(e^{-|\alpha_0+i\eta-\alpha_N|^2}\right)^2.
\end{equation}
We have $\omega_\text{t}t_\text{rep}\ll1$ so, expanding $|\alpha_0+i\eta-\alpha_N|^2$ to second order and the exponential to first order around zero, we find the expression in Eq.~(\ref{eq: error motion approx}).

%
%

\ack{We acknowledge financial support form the Spanish Government via the project PID2024-161371NB-C21 (MCIU/AEI/FEDER, EU) and project TSI-069100-2023-8 (Perte Chip-NextGenerationEU). E.T.,
acknowledges the Ram{\'o}n y Cajal (RYC2020-030060-I) research fellowship.}




\bibliography{bibliographyNJP}

\end{document}